\documentclass[twocolumn,amsmath,amssymb,aps,prb,superscriptaddress]{revtex4-2}
\usepackage{epsf}
\usepackage{graphicx}
\usepackage{sidecap}
\usepackage{soul}
\usepackage{array}
\usepackage{amsmath}
\usepackage{amssymb}
\usepackage{dcolumn}
\usepackage{epstopdf}
\usepackage{bm}
\usepackage{amsmath}
\usepackage{physics}

\usepackage{gensymb}
\usepackage{color}
\usepackage{hyperref}
\usepackage{soul}
\sethlcolor{green}
\usepackage{bbding}
\usepackage{setspace}
\hypersetup{
    colorlinks=true,
    citecolor=red,
    linkcolor=red,
    filecolor=blue,   
    urlcolor=blue,
}

\usepackage[normalem]{ulem}

\usepackage{graphicx,color}
\usepackage{amsfonts}
\usepackage[figuresright]{rotating}  
\usepackage{amssymb}
\usepackage{amsmath}

\usepackage{bbold}
\usepackage{wrapfig,lipsum,booktabs}
\usepackage{mathtools}
\usepackage{psfrag}
\usepackage{subfigure}
\usepackage{multirow}
\usepackage{tabularx}
\usepackage{textcomp}
\usepackage{bm}
\usepackage{hyperref}
\usepackage{relsize}
\hypersetup{
 pdfnewwindow=true, colorlinks=true,
 linkcolor=blue, anchorcolor=blue,
 citecolor=blue, filecolor=blue,
 menucolor=blue, urlcolor=blue}

\def\beq{\begin{eqnarray}}
\def\eeq{\end{eqnarray}}

\usepackage[dvipsnames]{xcolor}

\def \beq {\begin{equation}}
\def \eeq {\end{equation}}
\begin{document}

\title{Electronic band structure of a nodal line semimetal candidate ErSbTe}

\author{Iftakhar~Bin~Elius}\affiliation {Department of Physics, University of Central Florida, Orlando, Florida 32816, USA}
\author{Nathan~Valadez}\affiliation {Department of Physics, University of Central Florida, Orlando, Florida 32816, USA}
\author{Dante~James}\affiliation {Department of Physics, University of Central Florida, Orlando, Florida 32816, USA}
\author{Sami~Elgalal}\affiliation {Institute of Low Temperature and Structure Research, Polish Academy of Sciences, 50-950 Wrocław, Poland}
\author{Grzegorz~Chajewski}\affiliation {Institute of Low Temperature and Structure Research, Polish Academy of Sciences, 50-950 Wrocław, Poland}     
\author{Tetiana~Romanova}\affiliation {Institute of Low Temperature and Structure Research, Polish Academy of Sciences, 50-950 Wrocław, Poland}
\author{Andrzej~Ptok}\affiliation {Institute of Nuclear Physics, Polish Academy of Sciences, W. E. Radzikowskiego 152, PL-31342 Krak\'ow, Poland}
\author{Dariusz~Kaczorowski}\affiliation {Institute of Low Temperature and Structure Research, Polish Academy of Sciences, 50-950 Wrocław, Poland}
\author{Madhab~Neupane} \email[Corresponding author: ]{madhab.neupane@ucf.edu}
\affiliation {Department of Physics, University of Central Florida, Orlando, Florida 32816, USA}

\begin{abstract}
The $Ln$SbTe ($Ln =$ lanthanides) family is well known for hosting a plethora of intriguing characteristics stemming from its crystalline symmetry, magnetic structure, 4$f$ electronic correlations and spin--orbit coupling (SOC) phenomena. In this paper, we have systematically studied the bulk electrical and thermodynamic properties and electronic structure of the nodal line semimetal candidate ErSbTe using angle-resolved photoemission spectroscopy (ARPES) corroborated with first principles based theoretical band structure calculations with and without considering the effect of SOC, a critical factor dictating the band degeneracy which depends on the choice of the $Ln$ atom. Corroborative temperature dependent susceptibility, electrical resistivity and thermodynamic measurements, coherently exhibit paramagnetic to antiferromagnetic phase transition approximately at $1.94$~K, and another sharp anomaly at $1.75$~K. The zero field cooled resistivity measurement does not show the characteristic hump-like feature in the other $Ln$SbTe materials. The electronic band structure of ErSbTe, exhibits a diamond shaped Fermi surface. Along the high symmetry direction $\overline{\Gamma}$--$\overline{\text{X}}$, electronic bands are projected to cross over the Fermi energy, necessitated by the nonsymmorphic symmetry of the system. The other crossing along this direction is gapped, which evolves along the momentum space reaching its maximum along the $\overline{\Gamma}$--$\overline{\text{M}}$ direction. 
\end{abstract}

\maketitle

\section{Introduction}

From the onset of uncovering topological insulators (TIs), the intriguing realm of topological quantum materials (TQMs) has been established as a burgeoning frontier in the field of physics~\cite{col_hasan,Qi_TI, MNreview}. These materials serve as platforms for numerous fundamental physical phenomena and are continually advancing through rigorous theoretical and experimental investigations. This progression has led to the identification of a diverse array of topological semimetals, including but not limited to Dirac~\cite{wang, Neupane2014, Yang2014} and Weyl semimetals~\cite{Xu2015, Weylding, Soluyanov2015, Sakhya_2023_weyl}, nodal line semimetals (NLSMs)~\cite{Fang_NLSM_2015, node_surface_liang, MNzrsis, bian2016topological, hosen_ZrSiX, nodal_loop, Armitagereview, weyl_nodal_loop}, and Dirac nodal arcs~\cite{Wu2016_nodal_arc, gyanendra_ti2te2p}. Within systems that maintain both time-reversal symmetry and inversion symmetry, the occurrence of band crossings can culminate in the formation of Dirac nodes. At these nodes, two doubly degenerate bands converge, forming a four-fold degenerate crossing point. The linear dispersion relation around these Dirac nodes engenders massless Dirac fermions as the low-energy excitations within the system. These TQMs can be classified or differentiated based on the dimensionality of their band interactions or crossings in momentum space~\cite{wang, ZK_liu}. The concept of zero-dimensional band contact points in Dirac or Weyl semimetals is extended to encompass higher-dimensional nodal lines or surfaces~\cite{Chiu_2014, Fang_NLSM_2015, MNzrsis, pelayo_tii_dirac_2024}. In NLSMs, the band interactions manifest as lines or closed loops, which are protected by additional symmetries such as mirror reflection, inversion, time-reversal, spin-rotation, or nonsymmorphic symmetries~\cite{Fang_NLSM_2015, MNzrsis, nonsymmomorphic_Yang, schoop2016dirac}.

\begin{figure*}%[htb]
	\centering
	\includegraphics[width= \linewidth]{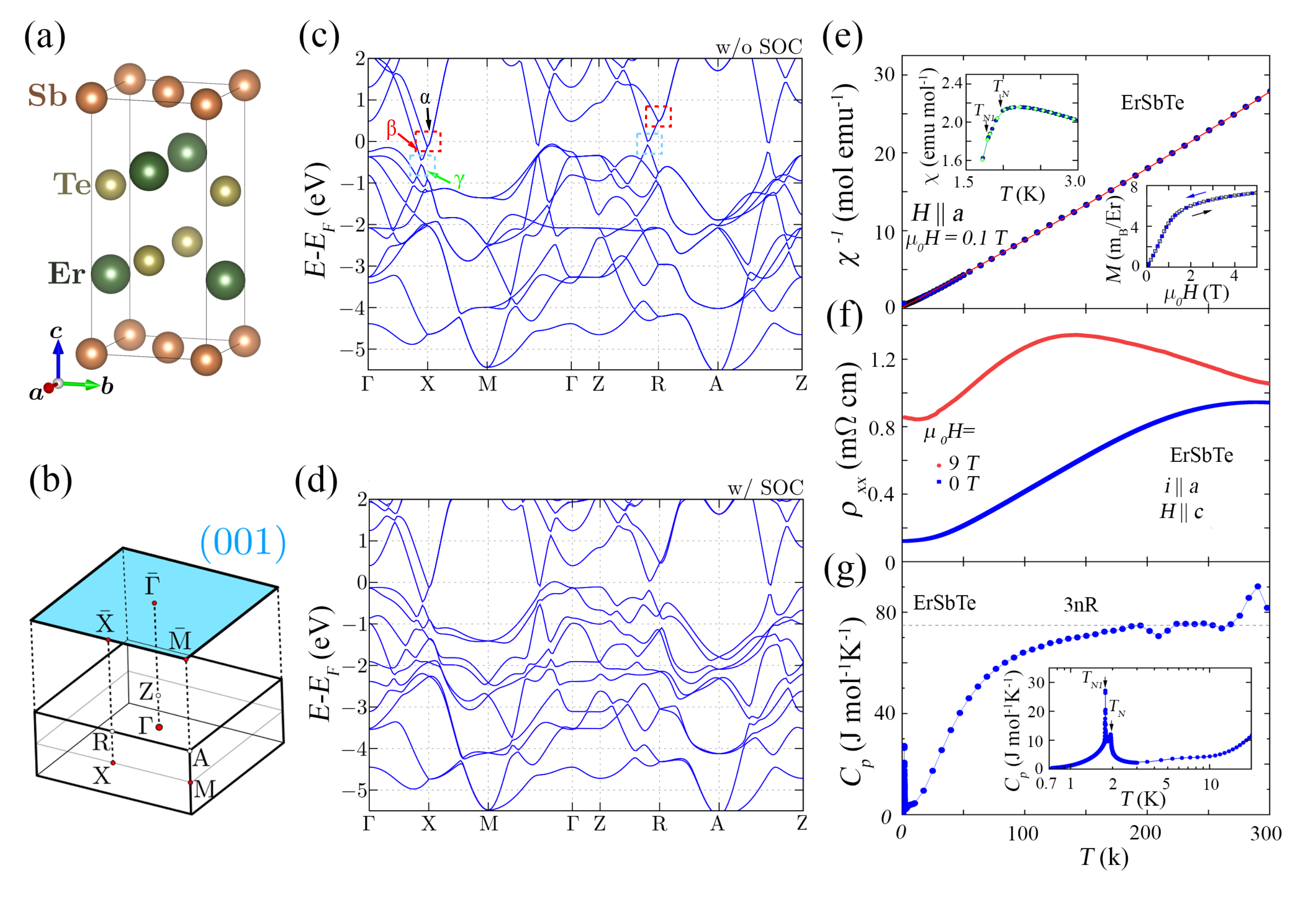}
	\caption
{\justifying Crystal structure, electronic band structure, thermodynamic, magnetic and transport properties of ErSbTe.
(a) Crystal structure of ErSbTe, containing 2D square planar Sb square-net and zigzag Er-Te layers. Electronic band structure (b) without and (c) with the spin--orbit coupling (SOC) along high symmetry directions of corresponding Brillouin zone presented on (b). Temperature dependencies of the (e) inverse magnetic susceptibility (f) electrical resistivity measured in magnetic fields of 0 and 9~T, and (g) specific heat of ErSbTe single crystals. The solid red line in panel (e) represents the Curie-Weiss fit. Upper inset in panel (e) depicts the magnetic susceptibility measured at the lowest temperatures measured in zero-field-cooled and field-cooled regimes (solid blue circles and open green diamonds, respectively); Lower inset in panel (e) presents the field dependence of the magnetization measured at 1.74~K with an increasing (open black squares) and decreasing (solid blue circles) magnetic field. The inset in panel (g) depicts the low-temperature specific heat data.
    \label{fig1}}
\end{figure*} 

\begin{figure*}%[htb]
	\centering
	\includegraphics[width= \linewidth]
{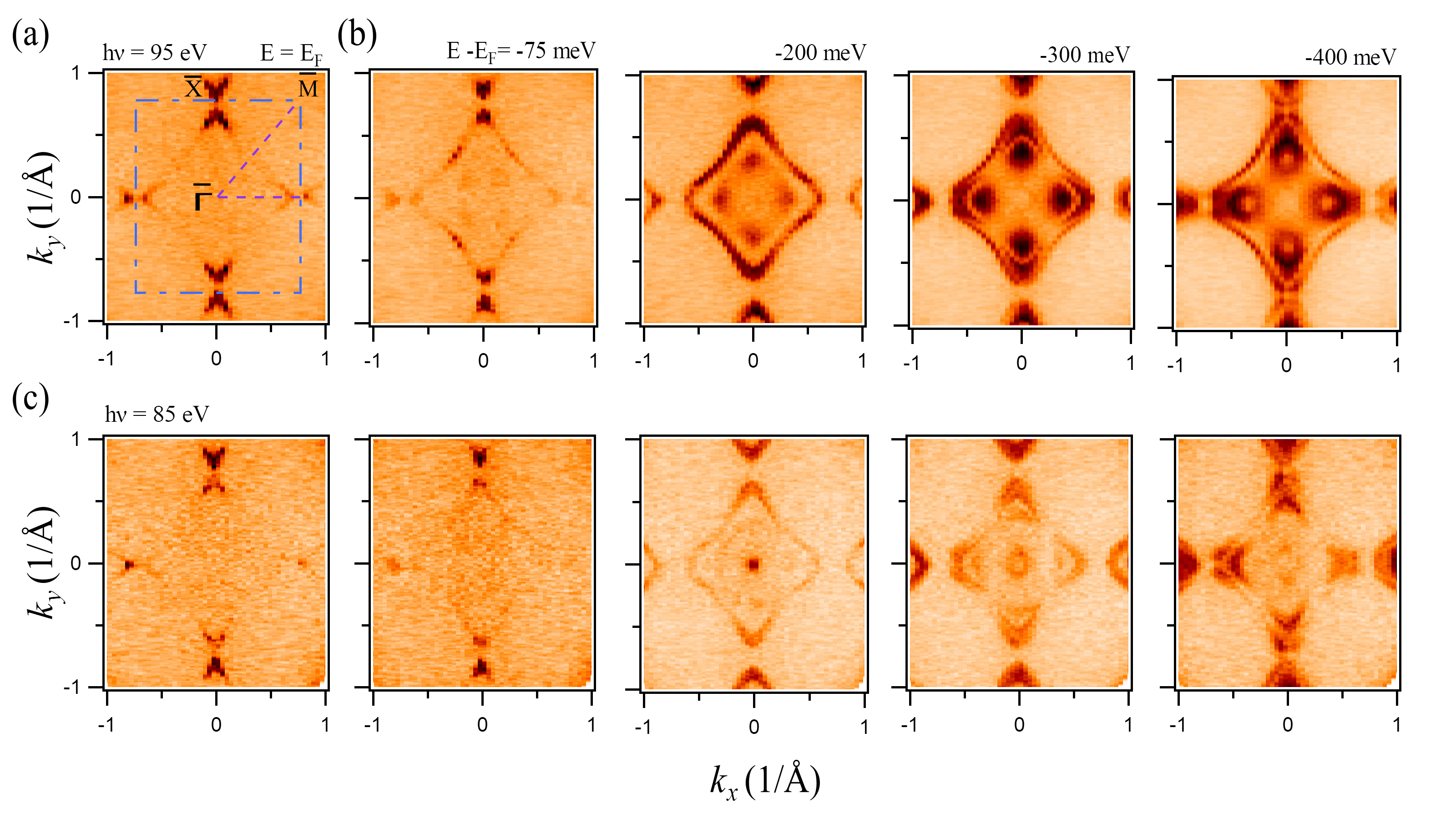}
	\caption{\justifying Fermi surface (FS) and constant energy contour (CEC) maps of ErSbTe. (a) Experimentally observed Fermi surface map and (b) CEC maps at different binding energies (mentioned at each subplot) of ErSbTe at an incident photon energy of 95~eV, the surface Brillouin zone is marked with a blue-dashed square and all the high symmetry points ($\overline{\Gamma}$, $\overline{\text{M}}$ and $\overline{\text{X}}$) are also indicated. (c) FS and CECs obtained from incident photon energy of 85~eV (measured at the respective binding energy positions). The ARPES measurements were performed at Stanford Synchrotron Radiation Lightsource (SSRL) endstation 5-2 at a temperature of 12~K.}
  \label{fig2}

\end{figure*} 
\indent Nonsymmorphic symmetry refers to a class of crystalline symmetry operations that integrate a point group transformation (e.g., rotation or reflection) with a fractional lattice translation, resulting in a symmetry element that is neither purely a point-group operation nor a simple translational shift~\cite{nonsymmomorphic_zhao}. The exploration of nonsymmorphic topological materials has been primarily guided by pivotal research in recent years, yet the repertoire of material families that feature non-accidental nodal lines in the absence of spin--orbit coupling (SOC) remains notably scarce~\cite{kane_theory_2016, schoop2016dirac, Schoop, nodaltheory}. Recent investigations have illuminated the potential for continuous Dirac nodal points, contingent upon the condition that 2D square motifs are configured into the hosting unit cell, thereby establishing glide symmetry~\cite{kane_theory_2016, Schoop, 
 GdSbTe_hosen,Shao2020}. These theoretical speculations were substantiated by the observation of nodal line topological phases initially in ZrSiS and subsequently in other $MZX$ ($M=$ Transition elements, $Z=$ Si, Ge, Sb, Sn, and $X=$ S, Se, Te) materials characterized by a PbFCl-type crystal structure~\cite{Neupane_16, schoop2016dirac}. When transition metals are substituted with rare earth elements, distinct topological characteristics are anticipated, influenced by the interactions between 4$f$ and conduction band electrons, alongside the inherent magnetism of 4$f$ states within the $Ln$SbTe ($Ln=$lanthanides) family. 
SOC is a quantum mechanical phenomenon that significantly influences the electronic properties of materials arising from the interaction between the electron spin and their orbital motion around the nucleus. This can lead to a variety of effects, such as the splitting of energy bands, which is crucial for the formation of topological insulators. Previous studies on $Ln$SbTe series, ARPES based studies corroborated by theoretical calculations established, a systematic effect of lanthanide incorporation on the SOC modification of the band structure. Investigations into lighter $Ln$-SbTe compounds, such as LaSbTe~\cite{La}, PrSbTe~\cite{prsbte_yuan_2024, PrSbTe}, NdSbTe~\cite{Nd111}, and SmSbTe~\cite{SmSbTe_sabin}, have consistently demonstrated the absence of a spin-orbit coupling (SOC)-induced band gap, at least within the limits of experimental resolution. Similarly, for intermediate $Ln$SbTe compounds like GdSbTe~\cite{GdSbTe_hosen} and TbSbTe~\cite{Tb111}, although SOC-induced band gaps were theoretically anticipated, they have not been empirically discerned in angle-resolved photoemission spectroscopy (ARPES) studies. Conversely, heavier counterparts such as HoSbTe and DySbTe are documented to exhibit fully gapped nodal line features~\cite{HoSbTe_arpes, valadez2025low}.\\
Moving forward to ErSbTe, the very existence of which has been very recently been reported in Ref.~\cite{plokhikh_Ln}, the study briefly explored the structural and magnetic properties of this compound. The scarcity of comprehensive studies on thermodynamic, electrical transport or electronic band structure of this compound, makes ErSbTe an alluring material to examine the effect of heavier $Ln$- atoms on the electronic structure of $Ln$SbTe series.

In this paper, we present the bulk electronic and thermodynamic transport properties of single crystal ErSbTe. The studies were followed by electronic band structure measurements on ErSbTe via ARPES, corroborated by the first principles calculations. Theoretical band structure calculations with and without considering SOC exhibit Dirac crossings parallel to the $\Gamma$--X high symmetry direction, which are parts of nodal lines extending along X--R. In this system, nonsymmorphic glide plane symmetry in combination with time reversal symmetry ($\mathcal{T}$) protects the out of plane nodal line along the X--R direction even with the effect of SOC considered. 

\begin{figure*}%[htb]
	\centering
	\includegraphics[width= \linewidth]
{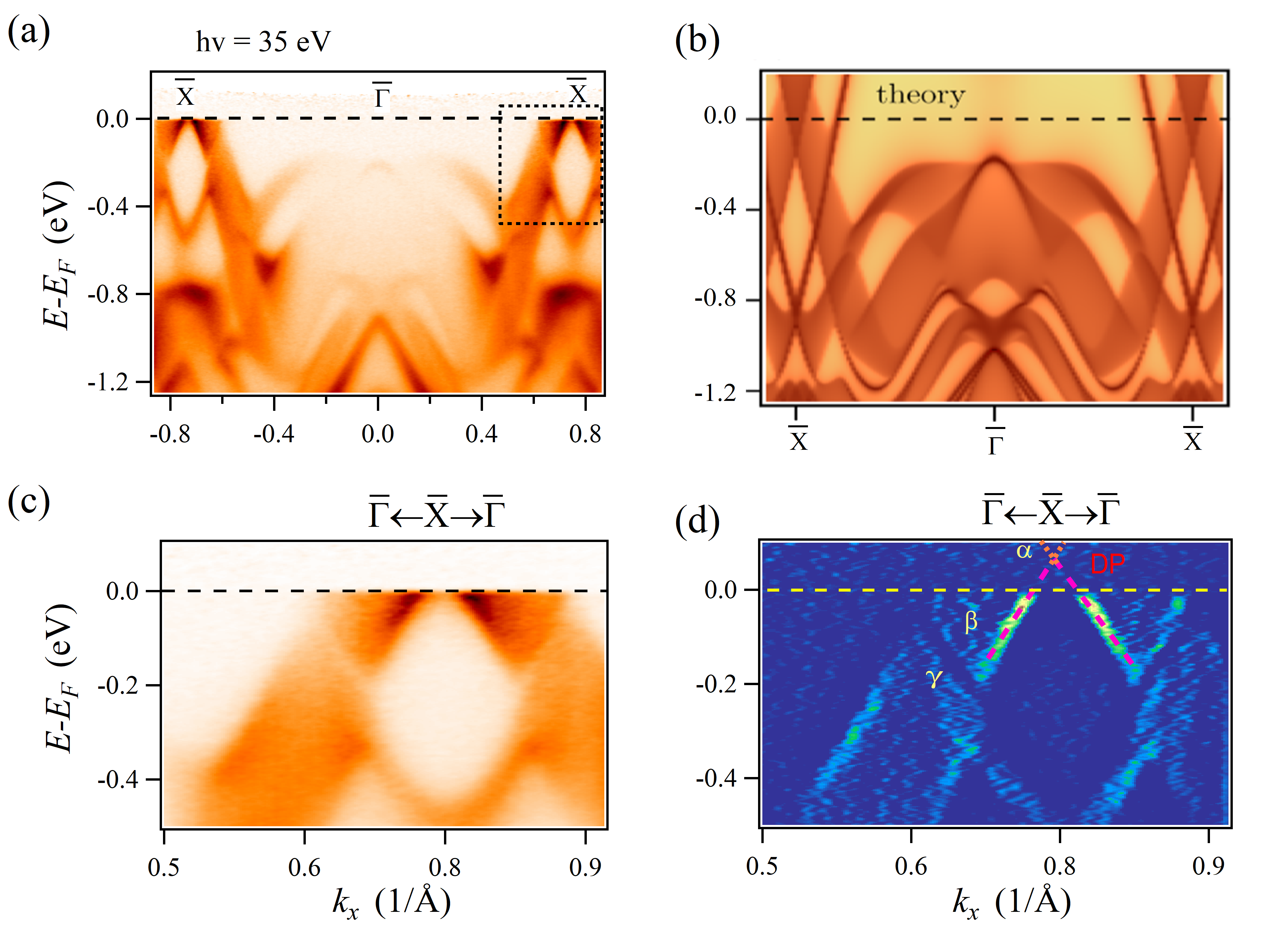}
	\caption{\justifying Electronic structure along the $\overline{\Gamma}$--$
\overline{\text{X}}$ direction. (a) Electronic dispersion map along $\overline{\text{X}}$--$\overline{\Gamma}$--$\overline{\text{X}}$ direction measured with incident photon energy of 35~eV, (b) the theoretically calculated surface projected band structure along $\overline{\Gamma}$--$\overline{\text{X}}$. (c) Magnified view of the dispersion map in the vicinity of $\overline{\text{X}}$ high symmetry point (indicated with the box in the panel (a)), (d) second derivative of Fig.~(c). The broken pink (orange) lines act as guide for the eyes indicating bands $\beta$ $(\alpha)$, converging to form the Dirac point. The ARPES dispersion maps were collected at beam line 5--2 in SSRL at a temperature of 12~K.}
  \label{gx}

\end{figure*} 

\section{Methods}
High-quality single crystals of ErSbTe were synthesized via the self-flux method. The quality of the crystals was confirmed by Laue diffraction, and their chemical composition was characterized using scanning electron microscopy (SEM) equipped with energy-dispersive X-ray spectroscopy (EDX). Further details on crystal growth and characterization procedures are provided in supplementary material (SM) art.~\textbf{A1}.\\
%\subsection*{Field induced thermodynamic and electrical transport measurements}
%\subsection*{Angle resolved photoemission spectroscopy}
\indent The electronic band structure measurements were performed at the Stanford Synchrotron Radiation Lightsource (SSRL) beamline endstation 5--2, equipped with a SCIENTA DA30L electron spectrometer. The energy resolution was set to be better than 20~meV and the angle resolution was better than 0.1\degree~for all the measurements. The samples were mounted on copper sample holders, then posts were attached to their upper surface using silver epoxy. Then they were transferred to the high vacuum ARPES chamber and cleaved $in~situ$ at a pressure better than $10^{-10}$~torr and measurements were performed at a temperature of 12~K (for details see SM art.~\textbf{A2}).\\ 
%\subsection*{Theoretical calculations}
\indent The {\it ab initio} calculations based on density functional thoery (DFT) were performed using the projector augmented-wave (PAW) potentials~\cite{blochl.94} implemented in the Vienna Ab initio Simulation Package ({\sc Vasp}) code~\cite{kresse.hafner.94,kresse.furthmuller.96,kresse.joubert.99}. Calculations are made within the generalized gradient approximation (GGA) in the Perdew, Burke, and Ernzerhof (PBE) parameterization~\cite{perdew.burke.96}.
The energy cutoff for the plane-wave expansion was set to $350$~eV, while $f$ electrons were treated as core states.
Optimizations of structural parameters (lattice constants and atomic positions) are performed in the primitive unit cell using the $15 \times 15 \times 7$ {\bf k}--point grid in the Monkhorst--Pack scheme~\cite{monkhorst.pack.76}.
As a break of the optimization loop, we take the condition with an energy difference of $10^{-6}$~eV and $10^{-8}$~eV for ionic and electronic degrees of freedom. 
The topological properties, as well as the electronic surface states, were studied using the tight binding model in the maximally localized Wannier orbitals basis~\cite{marzari.vanderbilt.97,souza.marzari.01}. 
This model was constructed from exact DFT calculations in a primitive unit cell (containing one formula unit), with $10 \times 10 \times 8$
 $\Gamma$-centered {\bf k}--point grid, using the {\sc Wannier90} software~
\cite{pizzi.vitale.20}.
As a starting projection we take $p$ and $d$ orbitals for Er, and $p$ orbitals for Sb and Te, what give 56 orbital 128 band tight binding model.
During calculations, the $f$ electrons of Er were treated as a core state.
The electronic surface states were calculated using the surface Green's function technique for a semi-infinite system~\cite{sancho.sancho.85}, implemented in {\sc WannierTools}~\cite{wu.zhang.18}.

\begin{figure*}%[ht]
	\centering
	\includegraphics
[width=\linewidth]{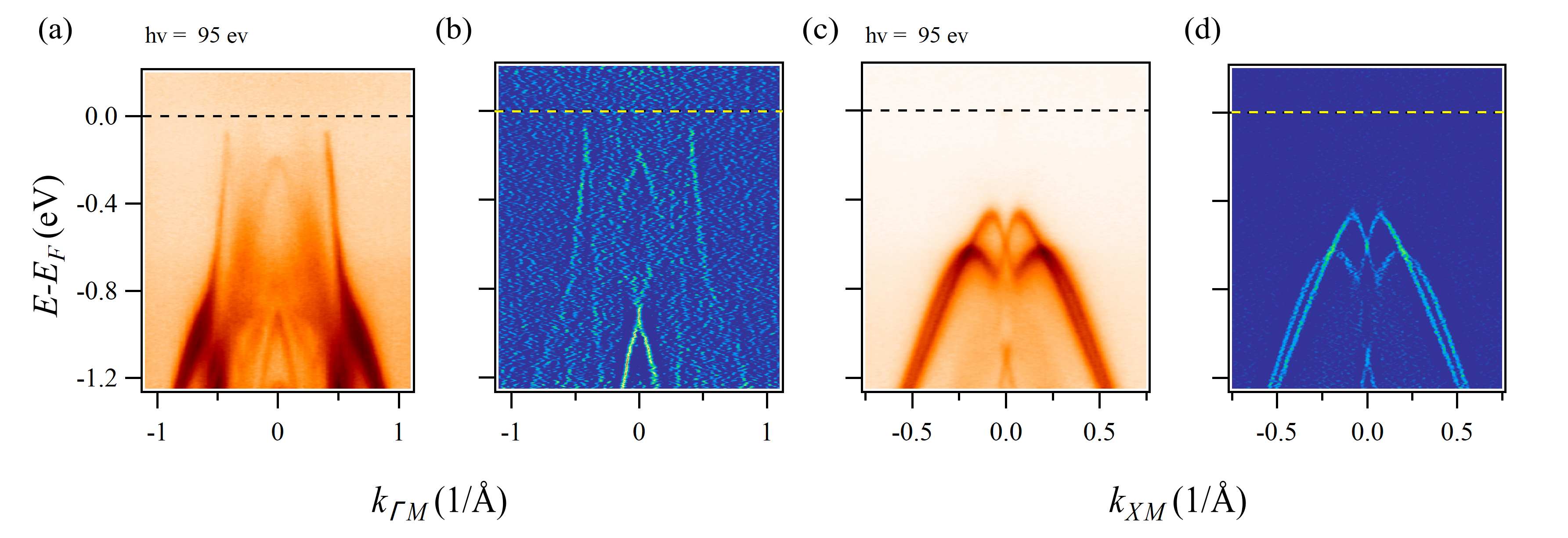}
	\caption{\justifying ARPES band dispersions along the $\overline{\Gamma}$--$\overline{\text{M}}$ and $\overline{\text{M}}$--$\overline{\text{X}}$ directions.
    (a) Cuts parallel to the  $\overline{\Gamma}$--$\overline{\text{M}}$ high symmetry direction. (b) second derivative of (a). (c) Band dispersion along the $\overline{\text{M}}$--$\overline{\text{X}}$ direction and (d) its second derivative. The measurements were performed at beamline 5--2 in SSRL at a temperature of 12~K.}
    \label{fig4}
\end{figure*}

\section{Results and Discussion}

ErSbTe crystallizes in the tetragonal $P4/nmm$ space group (No.~129), isostructural to other $Ln$SbTe materials. 
The calculated cell parameters are $a=b=4.266$~\AA, and $c= 9.184$~\AA. Structural investigations imply that, the Er atoms occupy the Wyckoff position at $2c$ ($\frac{1}{4},\frac{1}{4},0.2759$), Sb at $2a$ ($\frac{3}{4},\frac{1}{4},0$), and Te at $2c$ ($\frac{1}{4},\frac{1}{4},0.6244$), which are in good agreement with the previous report on ErSbTe~\cite{plokhikh_Ln}.
The crystal structure of ErSbTe is presented in Fig.~\ref{fig1}(a), the zig-zag Er-Te atomic chains are sandwitched by Sb square-nets. The calculated bulk band structure without and with consideration of SOC effect is presented in Figs.\ref{fig1}(b) and (c), respectively. 
%The positions of the nodal crossings are indicated with green boxes.
The bulk Brillouin zone with the high symmetry points and the directions are presented in Fig.~\ref{fig1}(d).
% \begin{figure*}[ht]
%	\centering
%	\includegraphics[width=\linewidth]{Fig_05_new.png}
%	\caption{\justifying Electronic structure parallel to $\overline{\Gamma}$--$\overline{\text{M}}$. ARPES spectrograph along $\overline{\Gamma}$--$\overline{\text{M}}$ direction measured with 95~eV of incident photon energy with (a) linear vertical (LV) and (b) linear horizontal (LH) polarization, (c) integrated image of both polarization, (d) second derivative of (c) and (e) Theoretically calculated surface projected spectrum along $\overline{\Gamma}$--$\overline{\text{M}}$. The ARPES measurements were performed at beamline 5-2 in SSRL at a temperature of 20~K. 
% }
%    \label{fig4}
%\end{figure*}
The main thermodynamic and electrical transport characteristics of the ErSbTe single crystals investigated in this study are presented in Fig.~\ref{fig1}(e-g). Fig.~\ref{fig1}(e) depicts the temperature dependence of the inverse magnetic susceptibility in a magnetic field of 0.1~T applied along the [100] crystallographic axis. Over a broad temperature range, the experimental data $\chi^{-1}$(T) follows the Curie--Weiss law, as indicated by the solid straight line in the main panel of Fig.~\ref{fig1}(e). A slight deviation from the linearity occurs only at temperatures below about 50~K. The displayed fit corresponds to the effective magnetic moment $\mu_{eff}$ = 9.77~$\mu_{B}$ and the paramagnetic Curie temperature $\theta_p$ = -1.6~K. The value of $\mu_{eff}$ is close to the theoretical value of 9.59~$\mu_{B}$, calculated for trivalent Er ion within the framework of Russel-Saunders coupling. In turn, the negative value of $\theta_p$ suggests weak antiferromagnetic exchange interactions. At the lowest temperatures, in the magnetic susceptibility of ErSbTe one can observe a slightly broadened maximum at about 2.2~K followed by a distinct decrease at $T_N$ = 1.95~K, indicating the onset of the long--range magnetic ordering (see the upper inset in Fig.~\ref{fig1}(e). Interestingly, this decrease seems to become even faster below $T_{N1}$ = 1.75~K. The lack of any bifurcation detected for data points collected in zero-field-cooled and field-cooled regimes points out the antiferromagnetic character of the magnetic order. As can be seen in the lower inset in Fig.~\ref{fig1}(e), in the ordered state, the magnetic field dependence of magnetization is typical for antiferromagnetic materials. At low fields, here up to about 1~T, $M$(H) shows nearly linear behavior, and in higher fields, it tends to saturate around 7~$\mu_B$/Er. Moreover, no hysteresis is observed for field-up and field-down sweep measurements.\\
\indent The temperature dependencies of the electrical resistivity of ErSbTe measured in 0 and 9~T magnetic fields are displayed in Fig.~\ref{fig1}(f). Surprisingly, the zero-field $\rho$(T) data does not show the semimetallic-like behavior with a broad humplike feature around $\sim$200~K, characteristic for $Ln$SbTe materials~\cite{Tb111, HoSBTe_yang}. Instead, it exhibits a metallic-like character, resembling that observed for ZrSiS \cite{Pressure_van} and numerous other materials adopting ZrSiS-type crystal structure (see e.g. Refs.~\cite{hosen_ZrSiX, ZnXSb_ciesielski, yang_zrsis_synthesis}). In turn, upon a magnetic field of 9~T applied parallel to the c-axis, the shape of $\rho$(T) is drastically changed and reminiscent of those of other $Ln$SbTe compounds. One may also note a clear difference in the low-temperature behavior of resistivity measured in 0 and 9~T. While in zero-field $\rho$ saturates at about 0.12~m$\Omega$cm, in a field of 9~T, it increases with decreasing temperature. However, because the resistivity measurements were only performed down to T = 2~K, no distinct signatures of the magnetic ordering were observed in our $\rho$(T) data.\\
\indent Fig.~\ref{fig1}(g) presents the temperature dependence of the specific heat. At temperatures above 200~K, $C_p$(T) levels off at about 75~J/(mol K)--a value nearly perfectly corresponding to the Dulong-Petit limit, $3nR$, where $n$ represents the number of atoms in the formula unit and R is the universal gas constant. The irregularities observed beyond 210~K are likely due to the use of Apiezon N grease as a thermal coupling medium~\cite{schnelle_apiezon}. Below 10~K, the compound exhibits a broad Schottky-like anomaly, which can be attributed to the splitting of the energy states in the crystal electric field. At the lowest temperatures, the specific heat of ErSbTe is dominated by distinct anomalies at $T_N$ = 1.94~K and $T_{N1}$ = 1.77~K. The former signals the phase transition into the antiferromagnetically ordered state and corresponds to the drop observed in the magnetic properties measurements. The latter peak in $C_p$(T) is much sharper, which suggests a first-order phase transition. Presumably, it is related to some reconfiguration of the magnetic structure of the compound and points to its complex magnetic behavior. \\

\indent Subsequently, we investigate the electronic structure of ErSbTe. Fig.~\ref{fig2} depicts the Fermi surface (FS) and constant energy contours (CECs) at various binding energies, as captured in the ARPES spectrographs of an ErSbTe single crystal sample. These measurements were conducted using incident photon energies of $95$~eV (Fig.~\ref{fig2}(a) and (b)) and $85$~eV (Fig.~\ref{fig2}(c) and (d)), respectively. The FS resembles the characteristic diamond shaped nature of the broader ZrSiS type materials centering the $\Gamma$ high symmetry point. Unlike iso-structural materials like PrSbTe~\cite{prsbte_yuan_2024, PrSbTe} or NdSbTe~\cite{Nd111}, the FS does not show double sheet nature. Notably, the band signatures are prominent in the proximity of X point, despite being absent exactly at the X point and significantly absent in the $\Gamma$--M direction (midway along the sides of the diamond shape). The band features gradually emerge at higher binding energies, along this direction. A hole pocket appears from approximately $\sim 200$~meV binding energy surrounding the $\Gamma$ point. Moving forward to Fig.~\ref{gx}(c), ARPES dispersion map along $\overline{\text{X}}$--$\overline{\Gamma}$--$\overline{\text{X}}$ direction measured with incident photon energy of 35~eV is displayed, the theoretically calculated surface projected band spectrum is presented in Fig.~\ref{gx}(b), which is a fair reproduction of the experimental dispersion map along this direction. $Ln$SbTe type materials are known to host a Dirac crossing at the $X$ high symmetry point, enforced by a combination of non symmorphic glide plane symmetry ($\mathcal{\Tilde{M}}_z$) and time reversal symmetry ($\mathcal{T}$). Another Dirac like crossing of bulk band is facilitated by screw axes symmetry, $\Tilde{C}_{2\nu,~(\nu= x,y)}$ combined with inversion symmetry, $\mathcal{P}$ along $\Gamma-X$ direction (a schematic of the directions of the possible nodal lines are presented in the SM Fig.~\textbf{S1})~\cite{Schoop, La}. The latter one is positioned a bit away from the $\overline{\text{X}}$ point. A magnified view of ARPES map near the $\overline{\text{X}}$ point is presented in Fig.~\ref{gx}(c) taken with 35~eV incident photon energy, and a second derivative in Fig~\ref{gx}(d), for enhanced view of the band dispersions. Bands marked with pink (orange) broken lines (representing $\alpha$ and $\beta$) are projected to intersect over the Fermi energy to form the gapless-symmetry enforced Dirac crossing. The other crossing is visibly gapped across the intersection region, possibly due to the effect of SOC. The SOC induced gap in this system remains ever elusive, as in many of similar materials this gap was predicted by theoretical calculations, yet remained beyond the detectable limit of the probing instruments~\cite{Nd111, PrSbTe, Tb111}. In another cut near the $\overline{\text{X}}$ point, taken with 47~eV (see SM Fig.~\textbf{S3}) incident photon energy, exhibiting similar band features, establishing the ubiquity of the gap across the $k_z$ axis. The binding energy versus $k_z$ maps acquired at regions C1 and C2 in SM Fig.~\textbf{S4}(a) are presented in SM Fig.~\textbf{S4}(b) and (c), respectively. The stacked energy distribution curves (EDCs) measured at the region indicated by the green arrow in Fig.~\textbf{S4}(b) consistently demonstrate the presence of the energy gap \\

\indent Next, we move forward to the other high symmetry directions. Dispersion cut along the high symmetry direction $\overline{\Gamma}$--$\overline{\text{M}}$ measured with an incident photon energy of 95~eV (a combination of LV and LH polarized beams), is presented in Fig.~\ref{fig4}(a) (which corresponds to $k_z$=3.7$\pi$/c). The bands $\alpha$ and $\beta$ move over the Fermi energy along this direction, opening a gap of approximately 75~meV. To better visualize the band features, a second derivative (calculated using 2D curvature method) plot of Fig.~\ref{fig4}(a) is presented in Fig.~\ref{fig4}(b). This also does not hint signature of band within this gap. ARPES dispersion cuts measured with different incident photon energies (see SM Fig.~\textbf{S5}) ranging from 55$\sim$95~eV, consistently show this gapped nature of the bands. The evolution of the gap, along the momentum space, is tracked in the  SM Figs.~\textbf{S6}(b) and \textbf{S6}(d). Moving to Fig.~\ref{fig4}(c) and its second derivative in Fig.~\ref{fig4}(d), dispersion map along the $\overline{\text{M}}$--$\overline{\text{X}}$ direction, exhibits no signature of bands over 400~meV of binding energy, unlike the other $Ln$SbTe materials. Our ARPES data along $\overline{\Gamma}$--$\overline{\text{X}}$ exhibited the Dirac crossing at the $\overline{\text{X}}$ point, occurs over the Fermi energy.  In this direction, the band features below 400~meV resembles the band features of other $Ln$SbTe's. More cuts along $\overline{\text{M}}$--$\overline{\text{X}}$ direction, measured with various incident photon energy suggests existence of surface bands in this direction (for detailed analysis see SM Fig.~\textbf{S7}).\\

\section{Conclusion}
In this study, single crystal specimens of ErSbTe were synthesized, followed by comprehensive bulk temperature dependent electrical and thermodynamic assessments, along with ARPES for band structure analysis, which was underpinned by calculations derived from first principles. The susceptibility and specific heat measurements indicate two antiferromagnetic orderings below 1.94~K and 1.75~K respectively. The electrical transport curve exhibited magnetization dependent features, which are unique when compared to other $Ln$SbTe materials. First principles based calculations exhibit presence of a nonsymmorphic symmetry protected nodal line along X--R direction. Analysis of ARPES based results along the $\overline{\Gamma}$--$\overline{\text{X}}$ direction exhibit bands intersecting over the Fermi energy. A gap was observed in the $\overline{\Gamma}$--$\overline{\text{M}}$ direction, which is exclusively visible in the heavier $Ln$SbTe materials. The direction dependence of the band intersection was probed. Consequently, ErSbTe emerges as an exemplary material that illustrates an interplay of symmetry, topological properties, and the influence of SOC.\\

\vspace{2ex}

\section*{acknowledgments} 
\vspace{-0.1 cm}
M.N. acknowledges the support from the National Science Foundation (NSF) CAREER Award No.~DMR-1847962, and the NSF Partnerships for Research and Education in Materials (PREM) Grant No.~DMR-2424976. D.K. was supported by the National Science Centre (Poland) under research grant 2021/41/B/ST3/01141. A.P. acknowledges the support of the National Science Centre (NCN, Poland) under Projects No.~2021/43/B/ST3/02166. This research utilized resources of the Stanford synchrotron radiation lightsource (SSRL) BL5--2. We express our gratitude to Dr.~Donghui Lu and Dr.~Makoto Hashimoto Mo for providing valuable support with the beamline at the SSRL.

\end{document}